\documentclass[twocolumn,showpacs,prl,superscriptaddress,amsmath,amssymb]{revtex4-1}

\usepackage{multirow}
\usepackage{graphicx}
\usepackage{dcolumn}
\usepackage{bm}
\usepackage{soul}
\usepackage{color}
\usepackage{array}

\begin{document}

\preprint{APS/123-QED}

\title{Symmetry energy dependence of long timescale isospin transport }

\author{K. Stiefel}
\email{stiefel@nscl.msu.edu}
\affiliation{National Superconducting Cyclotron Laboratory, Michigan State University, East Lansing, Michigan 48824, USA}
\affiliation{Department of Chemistry, Michigan State University, East Lansing, Michigan 48824, USA}
\author{Z. Kohley}
\affiliation{National Superconducting Cyclotron Laboratory, Michigan State University, East Lansing, Michigan 48824, USA}
\affiliation{Department of Chemistry, Michigan State University, East Lansing, Michigan 48824, USA}
\author{R. T. deSouza}
\affiliation{Department of Chemistry and Center for Exploration of Energy and Matter, Indiana University, Bloomington, Indiana 47405, USA}
\author{S. Hudan}
\affiliation{Department of Chemistry and Center for Exploration of Energy and Matter, Indiana University, Bloomington, Indiana 47405, USA}
\author{K. Hammerton}
\affiliation{National Superconducting Cyclotron Laboratory, Michigan State University, East Lansing, Michigan 48824, USA}
\affiliation{Department of Chemistry, Michigan State University, East Lansing, Michigan 48824, USA}

\date{\today}

\begin{abstract}
Isospin transport occurring within dinuclear projectile-like fragments (PLFs) produced in heavy-ion collisions is explored as a probe of the nuclear symmetry energy.  Within the framework of the Constrained Molecular Dynamics model (CoMD), the existence of the long-lived dinuclear PLFs, for up to 800~fm/c, is observed.  It is demonstrated that changes in the $\langle N/Z \rangle$ of the two fragments resulting from the breakup of the dinuclear PLF is due to isospin transport.  The rate of the transport between the two fragments is shown to be dependent on the slope of the symmetry energy at saturation density.  Comparison of the CoMD calculations with experimental data establish that the evolution of $\langle N/Z  \rangle$ could be used to constrain the density dependence of the symmetry energy.
\end{abstract}


\maketitle

\par
\emph{Introduction.} The nuclear Equation of State (nEoS), which describes the fundamental properties of infinite nuclear matter, has been the motivation for a broad range of experimental studies. Of particular interest is how the nEoS evolves as a function of the neutron-to-proton ratio ($N/Z$) of nuclear matter, which is defined by the symmetry energy term of the nEoS~\cite{Li01,Tsang12,Fuc06}. The symmetry energy, which is often discussed in reference to its density dependence, $E_{sym}(\rho)$, represents the difference in binding energy between pure neutron matter and symmetric nuclear matter ($N=Z$). Constraining the form of the density dependence of the symmetry energy is critical for understanding the properties of asymmetric nuclear matter. For example, the masses, collective excitations, and neutron skin thicknesses of finite nuclei are dependent on $E_{sym}(\rho)$~\cite{Tsang12,CHEN05,Li08,Cent09,Roca11,Pat12,Moller12,Zhang13,Heb14,Tam14,Horo01}. Beyond the terrestrial laboratory, the symmetry energy has an essential role in the properties and evolution of neutron stars and core-collapse supernovae~\cite{Lattimer01,STEINER05,LI06,Jank07,Fischer14,Latt12,Stei12}.

\par
Constraints on $E_{sym}(\rho)$ are often reported based on the magnitude, $E_{sym}(\rho_{0})$, and the slope, $L = 3 \rho_{\circ} \frac{\partial E_{sym}(\rho)}{\partial \rho} \lvert _{\rho_{\circ}}$, of the symmetry energy at the saturation density of nuclear matter. While significant progress has been made towards defining the form of the symmetry energy~\cite{Roca11,Moller12,Li08,Tsang12,Kohley13,Latt12,Shetty07,Li01,Kohl10,Lat07,Sot12,Steiner_apj13}, a large uncertainty in the slope of the symmetry energy exists. Though compilations of the current constraints suggest a most likely value of $L$ within the range of 40 to 80~MeV, several measurements lie outside this range~\cite{TSANG09,Tsang12,Kohley14,Li13,Lattimer13,Vin14}.  Continued experimental and theoretical progress is required to reduce the uncertainty in our knowledge of density dependence of the symmetry energy, thus improving our understanding of neutron-rich nuclear matter.

\par
Recently, the binary breakup of an excited projectile-like fragment (PLF) formed in peripheral heavy-ion collisions at intermediate energies has been proposed as an environment to investigate isospin transport over long timescales~\cite{MCINTOSH10,Hudan12, Brown13, Hudan14}.  Collision of the projectile and target nuclei can result in the production of a transiently deformed and excited PLF~\cite{Boc00,Davin02,Colin03,Pia02,Fil05,Fil14}.  This PLF can subsequently undergo a binary decay (dynamical fission~\cite{Papa05,Fil05b,Rus10}) into a light and a heavy fragment.  Experimental measurements of the $\langle N/Z \rangle$ of the light fragment have been interpreted as isospin equilibration which occurs on timescales of 600-900~fm/c~\cite{Hudan14}.  This timescale is noteworthy as it is approximately six times longer than has been previously reported ~\cite{Steiner_05,TSANG04}.

\par
In this rapid communication, the experimental results are confronted with theoretical transport calculations within the framework of the Constrained Molecular Dynamics model (CoMD).  The mechanism for the formation of the dinuclear PLF and long timescale isospin transport is confirmed within the CoMD model. Furthermore, the rate and magnitude of the long lived isospin transport is linked to the symmetry energy.  The CoMD results demonstrate that measurements of isospin transport within dinuclear PLFs could provide a novel probe of the symmetry energy.

\par
\emph{Experimental and theoretical details.} The CoMD calculations were motivated by the experimental results from the 45~MeV/A $^{64}$Zn~+~$^{64}$Zn reaction measured using the FIRST charged particle array at the Cyclotron Institute at Texas A$\&$M University~\cite{Brown13}. The FIRST array had angular coverage from $4.5^{\circ}~< \theta_{lab} < 27^{\circ}$ and provided isotopic identification for $Z < 8$ fragments and elemental identification up to the beam ($Z = 30$). The binary decay events were selected by requiring that each event contained heavy ($Z_{H}$) and light ($Z_{L}$) fragments with $Z_{H} > 11$ and $Z_{L} > 3$. In the following, all presented CoMD results have the same event selection (angular coverage and $Z_{H}, Z_{L}$ criteria) to select on binary decay events, providing a realistic comparison to the experimental results.  Specifically, the events with $Z_{H} >$ 11 and $Z_{L} = 4$ are selected for comparison between the theory and experiment as this maximizes the statistics.  To provide a consistent comparison with the experimental data, $^{8}$Be fragments produced within the CoMD calculation were assumed to decay into $2\alpha$ particles.

\par
The formation and decay of the dinuclear system requires a dynamical treatment of the heavy-ion collision. The CoMD model was chosen as it provides a dynamical description of the evolution of the many-body system. Each nucleon in the system is represented by a Gaussian wave-packet and is propagated according to the derived equations of motion and effective Skyrme interaction. Special care is given to constraining the equations of motion in order to respect the Pauli principle and provide conservation of total angular momentum. Additional details about the CoMD model (officially referred to as CoMD-II) can be found in Refs.~\cite{PAPA01,Papa05}. Within the CoMD framework, the 45~MeV/A $^{64}$Zn~+~$^{64}$Zn reaction was simulated for 1000~fm/c.  At each time-step, a coalescence procedure based on the position of each nucleon is applied to identify the fragments.  After the initial projectile-target collision, the PLF is identified.  The time evolution of the PLF is followed and if it undergoes a binary breakup, the resulting $Z_{H}$ and $Z_{L}$ fragments are identified.  The possible secondary decay of the heavy and light fragments was not included in the simulation.

\par
\begin{figure}
\includegraphics[width=0.4\textwidth]{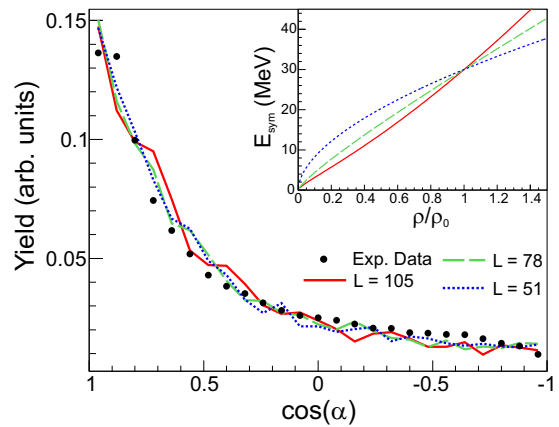}
\caption{\label{f:cosa} (Color online) Cos($\alpha$) distribution from the experiment (solid black circles) and CoMD simulation (lines) for events with $Z_{H} > 11$ and $Z_{L} = 4$. All results are normalized to a total integral of 1. The three different forms of the density dependence of the symmetry energy used within the CoMD calculations are shown in the insert. }
\end{figure}

The isospin-dependent part of the interaction used within CoMD can be varied to provide three different forms of the symmetry energy. The mean-field approximation of each parameterization is shown in the insert of Fig.~\ref{f:cosa} and is labeled based on the associated slope, $L$.  In all cases the magnitude of the symmetry energy at $\rho_{0}$ is 30~MeV. For each of the three forms of the symmetry energy, approximately 350,000 events were simulated over a triangular impact parameter range from 0 to 12~fm. Lastly, it is worth noting that the CoMD model has been successful in both describing complex heavy-ion collision dynamics over long timescales (including dynamical fission) and providing constraints on $E_{sym}(\rho)$ through comparison with experimental data~\cite{Kohley13,Papa07,Amo09,Card12}.

\par
\emph{Results and discussion.} In many cases an excited PLF with a dinuclear configuration can be produced following the nucleon exchange between the projectile and target nuclei. The binary decay of the dinuclear PLF into two fragments, designated $Z_{H}$ and $Z_{L}$, can be conveniently expressed in terms of their relative velocity, $v_{REL} = v_{H}-v_{L}$, and their center-of-mass velocity, $v_{CM}$. For a rotating dinuclear system, the primary experimental observable used to examine the dynamics of the reactions is the angle, $\alpha$, between $v_{CM}$ and  $v_{REL}$.  For cos($\alpha$) $> 0$, the lighter fragment ($Z_{L}$) is emitted ``backwards'' or towards the target with respect to the heavier fragment ($Z_{H}$).  In the case of cos($\alpha$) $< 0$, the lighter fragment is emitted forwards of the heavy fragment.  Thus, cos($\alpha$) measures the rotation of the dinuclear system.   A value of cos($\alpha$)~=~1 represents no rotation of the dinuclear system.  The cos($\alpha$) distribution of the $Z_{L} = 4$ fragments from the experiment is presented in Fig.~\ref{f:cosa}. The CoMD results, shown as the lines in Fig.~\ref{f:cosa}, provide an excellent description of the experimental $cos(\alpha)$ distribution.

\par
The rotation of the system described by cos($\alpha$) provides a measure of the lifetime of the dinuclear system.  Thus, the large yield observed for cos($\alpha$) $> 0.5$ indicates a preference for breakup on a short timescale relative to the rotation of the PLF~\cite{Brown13}.  Events with cos($\alpha$) $< 0$ are related to a longer rotation of the dinuclear system in which the $Z_{L}$ fragment is emitted forward relative to the $Z_{H}$ fragment.  The good agreement between the calculations and experimental data shown in Fig.~\ref{f:cosa} indicates that the fundamental dynamics of the reaction are reproduced. It is important to note that this agreement between calculations and data is independent of $L$, indicating that the dynamics of the dinuclear breakup are not sensitive to the symmetry energy.

\par
\begin{figure}
\includegraphics[width=0.4\textwidth]{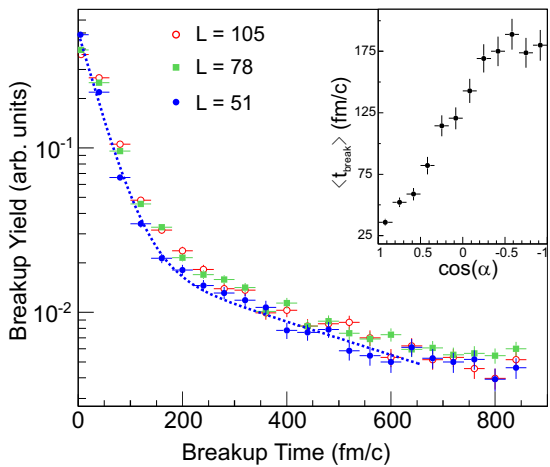}
\caption{\label{f:bt} (Color online) Yield of dinuclear breakup events with $Z_{H}>11$ and $Z_{L} = 4$ as a function of the time of breakup in the CoMD simulation. A two-component exponential decay fit to the $L = 51$~MeV results is shown by the dotted line. The correlation between the average breakup time ($t_{break}$) of the dinuclear PLF and cos($\alpha$) is shown in the insert. }
\end{figure}

The distribution of breakup times of the dinuclear PLFs from CoMD is shown in Fig.~\ref{f:bt}.  Separation of the projectile-like and target-like fragments corresponds to a breakup time of 0~fm/c, which occurs approximately 100-150~fm/c after initial contact of the projectile and target nuclei.  While dinuclear PLFs are observed to survive for up to $\sim$800~fm/c, the most probable breakup time is considerably shorter.  The prediction of long timescales is in reasonable agreement with the experimental results which reported dinuclear PLFs surviving for 600-900~fm/c based on estimating the rotational angular momentum for the system, which is discussed in more detail below~\cite{Brown13, Hudan12, Hudan14}.  For different values of $L$, slight differences are observed. To better quantify the distributions and the $L$ dependence, each distribution was fit with a two-component exponential decay, as shown by the dotted line for $L = 51$~MeV.  The extracted lifetimes for each component (short and long) are presented in Table~\ref{table}. The average lifetime of the dinuclear system is sensitive to $L$ with a decreased $L$ corresponding to a decrease in the average lifetime. A similar trend was observed by the CHIMERA collaboration examining incomplete fusion (or breakup) in low-energy reactions~\cite{Amo09}.

\par
The relationship between the average breakup time and cos($\alpha$) angle was also extracted from CoMD, as shown in the insert of Fig.~\ref{f:bt}. A clear correlation is present, verifying that the cos($\alpha$) observable is sensitive to the lifetime of the dinuclear PLF.  The relatively short range of the average breakup time from 25 to 175~fm/c is due to the dominance of short timescales as evident in Fig.~\ref{f:bt}.  This correlation between average breakup time and cos($\alpha$), which was hypothesized in the experimental analysis~\cite{Brown13, Hudan12, Hudan14}, is borne out by the present calculations.  However, for cos($\alpha$)~$< -0.25$ the correlation with $t_{break}$ is diminished within the CoMD simulation, indicating that events with cos($\alpha$) near $-1$ will not provide a direct link to the longest timescale events.  While the angular momentum of the dinuclear systems extracted from the experimental analysis, $J = 6 \pm 1 \hbar$~\cite{Brown13}, is in reasonable agreement with the average angular momentum obtained from CoMD, $J = 8.6 \hbar$, the average rotational frequency estimated from the insert of Fig.~\ref{f:bt} is roughly 10 times larger than that extracted from the experimental analysis~\cite{Brown13}.  This, again, indicates that within the CoMD description the cos($\alpha$) observable is most sensitive to the shorter time scales due to the faster rotation of the dinuclear system.

\begin{table}
\begin{center}
\caption{Average lifetimes extracted from the two-component fit of Fig.~\ref{f:bt} for the different forms of $E_{sym}(\rho)$.}
\begin{tabular}{c c c}
\hline
\hline
$L$ (MeV) &$\langle\tau_{short}\rangle$ (fm/c) &$\langle\tau_{long}\rangle$ (fm/c)\\
\hline
51 &38 &429 \\
78 &52 &593 \\
105 &57 &711 \\
\hline
\hline
\end{tabular}
\label{table}
\end{center}
\end{table}

\begin{figure}
\includegraphics[width=0.4\textwidth]{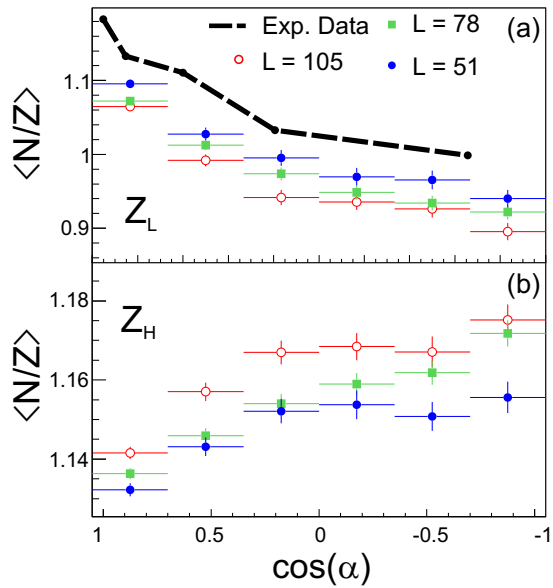}
\caption{\label{f:nz_cos} (Color online) Relationships between the $\langle N/Z \rangle$ and cos($\alpha$) for the $Z_{L}$ fragment and the $Z_{H}$ fragment extracted from the CoMD simulation are shown in panels (a) and (b), respectively. Experimental data for the $Z_{L}$ fragment is included for comparison. }
\end{figure}

\par
One intriguing aspect of the long-lived dinuclear systems is the possibility to observe and study isospin transport on timescales not previously thought possible.  The experimental evidence of this isospin transport is shown in Fig.~\ref{f:nz_cos}(a), as the dashed black line, where the $\langle N/Z \rangle$ of the light fragment ($Z_{L}=4$) is plotted as a function of cos($\alpha$).  As the rotation angle increases, the $\langle N/Z \rangle$ of the $Z_{L}$ fragment decreases.  Since the value of $Z_{L}$ is fixed at $Z = 4$, the change in $\langle N/Z \rangle$ corresponds to a flow of neutrons out of the $Z_{L}$ fragment. This neutron flow out of the $Z_{L}$ fragment has been interpreted as the transport of neutrons from the $Z_{L}$ fragment to the $Z_{H}$ fragment as the system rotates.  The CoMD results are also shown and while the $\langle N/Z \rangle$ of the $Z_{L}$ fragment is underpredicted, the trend of the experimental data is well reproduced.   Experimentally it has been determined that the larger the neutron-richness of the target, the larger the initial neutron-richness of the $Z_{L}$ fragment~\cite{Brown13}.    We speculate that the underprediction by the model is due to its failure to properly describe the projectile-target interaction which produces the dinuclear PLF.

\par
If the observed decrease in $\langle N/Z \rangle$ with rotation for $Z_{L}$ is due to nucleon transport to $Z_{H}$, then the $\langle N/Z \rangle$ of the $Z_{H}$ fragment should increase with rotation.  As evident in Fig.~\ref{f:nz_cos}(b), this expectation is met in the CoMD calculations, indicating that isospin transport occurs between $Z_{H}$ and $Z_{L}$ as the dinuclear system rotates.  Unfortunately, as the experimental measurement did not have isotopic identification of the $Z_{H}$ fragment, this isospin transport could not be experimentally determined.  It is important to note that the smaller change in magnitude of the $\langle N/Z \rangle$ of the $Z_{H}$ fragment compared to the $Z_{L}$ fragment is due to the larger $Z$ of the $Z_{H}$ fragment.

\par
Within the CoMD framework, the $\langle N/Z \rangle$ of both the $Z_{H}$ and $Z_{L}$ fragments as a function of cos($\alpha$) are observed to be sensitive to the form of the symmetry energy (Fig.~\ref{f:nz_cos}). The softest form of the symmetry energy ($L = 51$~MeV) produces the most neutron-rich $Z_{L}$ fragments in comparison to the stiffer forms of the symmetry energy.  This difference in the $\langle N/Z \rangle$ for different $L$ exists for cos($\alpha$) $\approx$ 1, i.e. the shortest times, and persists as the dinuclear system rotates.  Complementary behavior is observed for the $Z_{H}$ fragment with the softest symmetry energy producing the smallest $\langle N/Z \rangle$.  The CoMD calculations establish that the $\langle N/Z \rangle$ of the $Z_{L}$ and $Z_{H}$ fragments as a function of cos($\alpha$) is a sensitive observable to the form of the symmetry energy.

\begin{figure}
\includegraphics[width=0.35\textwidth]{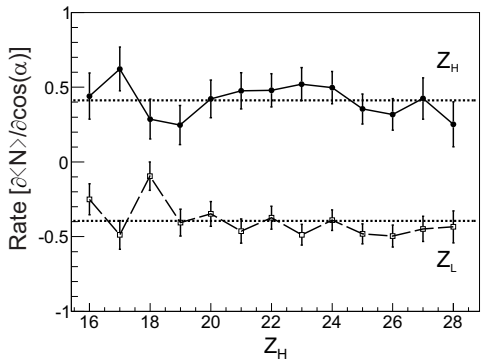}
\caption{\label{f:rate} The rate of change in the neutron number of the $Z_{L} = 4$ and $Z_{H}$ fragments per cos($\alpha$).  The rate is shown as function of the $Z$ of the heavy fragment.  The horizontal lines are fits to the data used to extract average rate of neutron transfer.}
\end{figure}

\par
Although Fig.~\ref{f:nz_cos} clearly indicates that isospin transport between $Z_{H}$ and $Z_{L}$ is occurring, it remains to be determined if isospin transport is the dominant reason for the changes in $\langle N/Z \rangle$.  For example, light particle emission could also alter the $\langle N/Z \rangle$. In Fig.~\ref{f:rate} the neutron transport between the fragments is more quantitatively investigated.  Examination of $\langle N \rangle$ as a function of cos($\alpha$) for $Z_{L} = 4$ and $Z_{H} = 16-28$ fragment pairs allows one to extract the rate of change in $\langle N \rangle$ with respect to cos($\alpha$) for both $Z_{L}$ and $Z_{H}$.  As evident in Fig.~\ref{f:rate}, the $Z_{L}$ fragment loses on average 0.39$\pm$0.02 neutrons per cos($\alpha$) while the $Z_{H}$ fragment gains on average 0.41$\pm$0.03 neutrons per cos($\alpha$).  The fact that the rate of neutron exchange is equivalent proves that isospin transport dominates the evolution of the $\langle N/Z \rangle$ of the $Z_{L}$ and $Z_{H}$ fragments.

\begin{figure}
\includegraphics[width=0.36\textwidth]{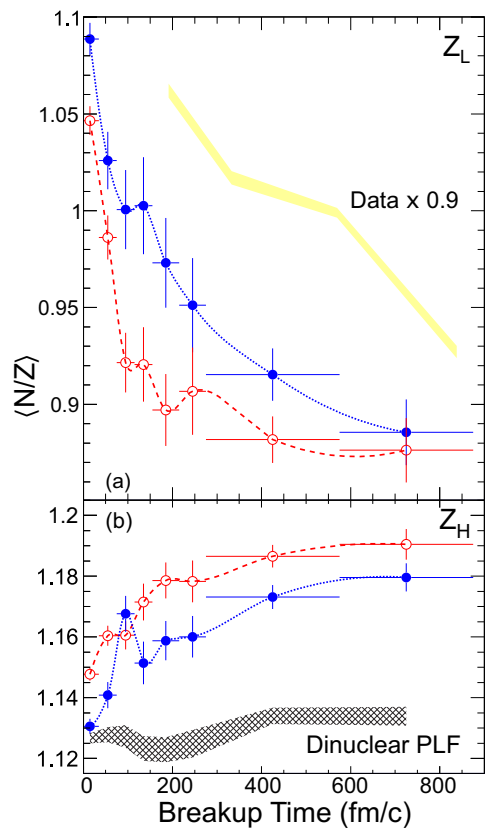}
\caption{\label{f:nz_time} (Color online) $\langle N/Z \rangle$ of the $Z_{L} = 4$ and the $Z_{H}$ fragments as a function of the breakup time are presented in panels (a) and (b), respectively. For simplicity, only $L = 51$~MeV (blue filled circles) and $L = 105$~MeV (red open circles) are shown. The experimental data (yellow region) for the $Z_{L}$ is scaled by 0.9.  The grey hatched area is the $\langle N/Z \rangle$ of the dinuclear PLF.  }
\end{figure}

\par
The dependence of $\langle N/Z \rangle$ of $Z_{L}$ and $Z_{H}$ on breakup time is presented in Fig.~\ref{f:nz_time}.  As can be expected from Fig.~\ref{f:nz_cos}, the $\langle N/Z \rangle$ of the $Z_{L}$ fragment decreases while that of the $Z_{H}$ increases with increasing breakup time.  It is remarkable that the $\langle N/Z \rangle$ of both the heavy and light fragment evolves for the longest breakup times observed, namely 800~fm/c.  Since it was established in Fig.~\ref{f:rate} that the change in $\langle N/Z \rangle$ of $Z_{H}$ is due to nucleon transport from the $Z_{L}$ fragment, we focus on the dependence of $\langle N/Z \rangle$ on breakup time for $Z_{L}$ where the change in $\langle N/Z \rangle$ is the largest.   Both the magnitude and rate of change of the $\langle N/Z \rangle$ are sensitive to the density dependence of the symmetry energy.  For the $Z_{L}$ fragment, the stiff symmetry energy ($L = 105$~MeV) exhibits a lower $\langle N/Z \rangle$ at $t_{break} = 0$~fm/c and a slightly faster decrease of $\langle N/Z \rangle$ with time as compared to the soft symmetry energy ($L = 51$~MeV), which has a higher initial $\langle N/Z \rangle$ and evolves more slowly.  For comparison, the experimental data for the $\langle N/Z \rangle$ of the $Z_{L}$ fragment as a function of breakup time is shown in Fig.~\ref{f:nz_time}(a)~\cite{Brown13}.  The experimental results are, of course, dependent on the estimated rotational frequency.  As discussed above, CoMD indicates a faster rotational frequency which would result in the experimental data (yellow region) shifting to lower breakup times in better agreement with the simulation.

\par
The $\langle N/Z \rangle$ of the dinuclear PLF, determined from the combination of the binary fragments, is shown in Fig.~\ref{f:nz_time}(b).  The relatively constant $\langle N/Z \rangle$ of the dinuclear PLF is consistent with the rate of neutron exchange discussed in Fig.~\ref{f:rate}, confirming that the changing compositions of the $Z_{L}$ and $Z_{H}$ fragments are dominated by the exchange of neutrons within the dinuclear system.

\par
The sensitivity of $\langle N/Z \rangle$ with breakup time to $L$ can be understood in terms of an initial isospin equilibration followed by a  competition of the dinuclear PLF to ``fuse'' (producing a single PLF) rather than break up into the $Z_{H}$ and $Z_{L}$ fragments~\cite{Amo09,Card12}. The dinuclear system is formed during the projectile-target interaction.  During this formation, the symmetry energy drives the initial $N/Z$ gradient of the dinuclear system towards equilibrium~\cite{TSANG04}.  The $\langle N/Z \rangle$ of $Z_{L}$ ($Z_{H}$) at $t_{break}=0$~fm/c is increased (decreased) for the soft $E_{sym}$, which brings them closer to isospin equilibrium, relative to the stiff $E_{sym}$. This initial sensitivity of the $\langle N/Z \rangle$ to $L$ is maintained throughout the evolution of the dinuclear system.
\par
Following the formation of the dinuclear PLF, the system then will either fuse (producing a single PLF) or undergo a binary breakup.  If the dinuclear PLF was to fuse, the $Z_{H}$ fragment would swallow up the $Z_{L}$ fragment.  Since we have examined binary breakup events with the requirement of $Z_{L}=4$, this fusion process is represented by the $Z_{H}$ fragment taking neutrons from the $Z_{L}$ fragment (evolution observed in Fig.~\ref{f:nz_time}).  Since the soft symmetry energy is more repulsive for neutron-rich systems at sub-saturation densities it will hinder the fusion of the dinuclear system relative to the stiff symmetry energy~\cite{Amo09,Card12}.  Thus, the stiff symmetry energy should allow the system to evolve towards fusion more easily (being less repulsive at low density), which would be characterized by an increased rate of neutron transfer from the $Z_{L}$ to the $Z_{H}$ fragment.  This interpretation is supported by Fig.~\ref{f:nz_time} which shows an increased rate of change in the $\langle N/Z \rangle$ for $L=105$~MeV relative to $L=51$~MeV.

\emph{Conclusions.} Microscopic transport calculations with the CoMD model confirm the physical picture that binary breakup of an excited projectile-like fragment produced in peripheral heavy-ion collisions at intermediate energies provides a useful probe of the density dependence of the symmetry energy.  The transport calculations were successful in reproducing the overall dynamical features of the binary decay.  A clear correlation is observed between the rotation angle of the dinuclear PLF and its lifetime providing a ``clock'' of the decay which is most sensitive for breakup times of $< 200$~fm/c.  The experimentally observed dependence of $\langle N/Z \rangle$ on rotation angle for the light fragment ($Z_{L}$) is qualitatively described by the model, however CoMD suggests that the rotational frequency of the dinuclear PLF is significantly faster than originally estimated~\cite{Brown13}.  Within the context of the model, the change in $\langle N/Z \rangle$ of light and heavy fragments with rotation is clearly dominated by nucleon exchange between the two fragments.  The isospin transport persists for the longest times observed, namely 800~fm/c.   Within the CoMD framework, the evolution of the $\langle N/Z \rangle$ with breakup time exhibits a definite dependence on the slope of the symmetry energy, providing a novel experimental probe of the nuclear Equation of State.

\emph{Acknowledgments.}  The authors thank M. Papa and A. Bonasera for use of the CoMD code.  This work is supported by the National Science Foundation under Grant No. PHY-1102511 (Michigan State University).  This material is based upon work supported by the U.S. Department of Energy Office of Science, Office of Nuclear Physics under Award Number DE-FG02-88ER-40404 (Indiana University).


\begin{thebibliography}{50}%
\makeatletter
\providecommand \@ifxundefined [1]{%
 \@ifx{#1\undefined}
}%
\providecommand \@ifnum [1]{%
 \ifnum #1\expandafter \@firstoftwo
 \else \expandafter \@secondoftwo
 \fi
}%
\providecommand \@ifx [1]{%
 \ifx #1\expandafter \@firstoftwo
 \else \expandafter \@secondoftwo
 \fi
}%
\providecommand \natexlab [1]{#1}%
\providecommand \enquote  [1]{``#1''}%
\providecommand \bibnamefont  [1]{#1}%
\providecommand \bibfnamefont [1]{#1}%
\providecommand \citenamefont [1]{#1}%
\providecommand \href@noop [0]{\@secondoftwo}%
\providecommand \href [0]{\begingroup \@sanitize@url \@href}%
\providecommand \@href[1]{\@@startlink{#1}\@@href}%
\providecommand \@@href[1]{\endgroup#1\@@endlink}%
\providecommand \@sanitize@url [0]{\catcode `\\12\catcode `\$12\catcode
  `\&12\catcode `\#12\catcode `\^12\catcode `\_12\catcode `\%12\relax}%
\providecommand \@@startlink[1]{}%
\providecommand \@@endlink[0]{}%
\providecommand \url  [0]{\begingroup\@sanitize@url \@url }%
\providecommand \@url [1]{\endgroup\@href {#1}{\urlprefix }}%
\providecommand \urlprefix  [0]{URL }%
\providecommand \Eprint [0]{\href }%
\providecommand \doibase [0]{http://dx.doi.org/}%
\providecommand \selectlanguage [0]{\@gobble}%
\providecommand \bibinfo  [0]{\@secondoftwo}%
\providecommand \bibfield  [0]{\@secondoftwo}%
\providecommand \translation [1]{[#1]}%
\providecommand \BibitemOpen [0]{}%
\providecommand \bibitemStop [0]{}%
\providecommand \bibitemNoStop [0]{.\EOS\space}%
\providecommand \EOS [0]{\spacefactor3000\relax}%
\providecommand \BibitemShut  [1]{\csname bibitem#1\endcsname}%
\let\auto@bib@innerbib\@empty
\bibitem [{\citenamefont {Li}\ and\ \citenamefont {Schroder}(2001)}]{Li01}%
  \BibitemOpen
  \bibinfo {editor} {\bibfnamefont {B.~A.}\ \bibnamefont {Li}}\ and\ \bibinfo
  {editor} {\bibfnamefont {U.}~\bibnamefont {Schroder}},\ eds.,\ \href@noop {}
  {\emph {\bibinfo {title} {Isospin Physics in Heavy-Ion Collisions at
  Intermediate Energies}}}\ (\bibinfo  {publisher} {NOVA Science},\ \bibinfo
  {year} {2001})\BibitemShut {NoStop}%
\bibitem [{\citenamefont {Tsang}\ \emph {et~al.}(2012)\citenamefont {Tsang}
  \emph {et~al.}}]{Tsang12}%
  \BibitemOpen
  \bibfield  {author} {\bibinfo {author} {\bibfnamefont {M.~B.}\ \bibnamefont
  {Tsang}} \emph {et~al.},\ }\href@noop {} {\bibfield  {journal} {\bibinfo
  {journal} {Phys. Rev. C}\ }\textbf {\bibinfo {volume} {86}},\ \bibinfo
  {pages} {015803} (\bibinfo {year} {2012})}\BibitemShut {NoStop}%
\bibitem [{\citenamefont {Fuchs}\ and\ \citenamefont {Wolter}(2006)}]{Fuc06}%
  \BibitemOpen
  \bibfield  {author} {\bibinfo {author} {\bibfnamefont {C.}~\bibnamefont
  {Fuchs}}\ and\ \bibinfo {author} {\bibfnamefont {H.~H.}\ \bibnamefont
  {Wolter}},\ }\href@noop {} {\bibfield  {journal} {\bibinfo  {journal} {Eur.
  Phys. J. A}\ }\textbf {\bibinfo {volume} {30}},\ \bibinfo {pages} {5}
  (\bibinfo {year} {2006})}\BibitemShut {NoStop}%
\bibitem [{\citenamefont {Chen}\ \emph {et~al.}(2005)\citenamefont {Chen},
  \citenamefont {Ko},\ and\ \citenamefont {Li}}]{CHEN05}%
  \BibitemOpen
  \bibfield  {author} {\bibinfo {author} {\bibfnamefont {L.~W.}\ \bibnamefont
  {Chen}}, \bibinfo {author} {\bibfnamefont {C.~M.}\ \bibnamefont {Ko}}, \ and\
  \bibinfo {author} {\bibfnamefont {B.~A.}\ \bibnamefont {Li}},\ }\href@noop {}
  {\bibfield  {journal} {\bibinfo  {journal} {Phys. Rev. C}\ }\textbf {\bibinfo
  {volume} {72}},\ \bibinfo {pages} {064309} (\bibinfo {year}
  {2005})}\BibitemShut {NoStop}%
\bibitem [{\citenamefont {Li}\ \emph {et~al.}(2008)\citenamefont {Li},
  \citenamefont {Chen},\ and\ \citenamefont {Ko}}]{Li08}%
  \BibitemOpen
  \bibfield  {author} {\bibinfo {author} {\bibfnamefont {B.~A.}\ \bibnamefont
  {Li}}, \bibinfo {author} {\bibfnamefont {L.~W.}\ \bibnamefont {Chen}}, \ and\
  \bibinfo {author} {\bibfnamefont {C.~M.}\ \bibnamefont {Ko}},\ }\href@noop {}
  {\bibfield  {journal} {\bibinfo  {journal} {Phys. Rep.}\ }\textbf {\bibinfo
  {volume} {464}},\ \bibinfo {pages} {113} (\bibinfo {year}
  {2008})}\BibitemShut {NoStop}%
\bibitem [{\citenamefont {Centelles}\ \emph {et~al.}(2009)\citenamefont
  {Centelles}, \citenamefont {Roca-Maza}, \citenamefont {Vinas},\ and\
  \citenamefont {Warda}}]{Cent09}%
  \BibitemOpen
  \bibfield  {author} {\bibinfo {author} {\bibfnamefont {M.}~\bibnamefont
  {Centelles}}, \bibinfo {author} {\bibfnamefont {X.}~\bibnamefont
  {Roca-Maza}}, \bibinfo {author} {\bibfnamefont {X.}~\bibnamefont {Vinas}}, \
  and\ \bibinfo {author} {\bibfnamefont {M.}~\bibnamefont {Warda}},\
  }\href@noop {} {\bibfield  {journal} {\bibinfo  {journal} {Phys. Rev. Lett.}\
  }\textbf {\bibinfo {volume} {102}},\ \bibinfo {pages} {122502} (\bibinfo
  {year} {2009})}\BibitemShut {NoStop}%
\bibitem [{\citenamefont {Roca-Maza}\ \emph {et~al.}(2011)\citenamefont
  {Roca-Maza}, \citenamefont {Centelles}, \citenamefont {Vinas},\ and\
  \citenamefont {Warda}}]{Roca11}%
  \BibitemOpen
  \bibfield  {author} {\bibinfo {author} {\bibfnamefont {X.}~\bibnamefont
  {Roca-Maza}}, \bibinfo {author} {\bibfnamefont {M.}~\bibnamefont
  {Centelles}}, \bibinfo {author} {\bibfnamefont {X.}~\bibnamefont {Vinas}}, \
  and\ \bibinfo {author} {\bibfnamefont {M.}~\bibnamefont {Warda}},\
  }\href@noop {} {\bibfield  {journal} {\bibinfo  {journal} {Phys. Rev. Lett.}\
  }\textbf {\bibinfo {volume} {106}},\ \bibinfo {pages} {252501} (\bibinfo
  {year} {2011})}\BibitemShut {NoStop}%
\bibitem [{\citenamefont {Patel}\ \emph {et~al.}(2012)\citenamefont {Patel}
  \emph {et~al.}}]{Pat12}%
  \BibitemOpen
  \bibfield  {author} {\bibinfo {author} {\bibfnamefont {D.}~\bibnamefont
  {Patel}} \emph {et~al.},\ }\href@noop {} {\bibfield  {journal} {\bibinfo
  {journal} {Phys. Lett. B}\ }\textbf {\bibinfo {volume} {718}},\ \bibinfo
  {pages} {447} (\bibinfo {year} {2012})}\BibitemShut {NoStop}%
\bibitem [{\citenamefont {Moller}\ \emph {et~al.}(2012)\citenamefont {Moller},
  \citenamefont {Myers}, \citenamefont {Sagawa},\ and\ \citenamefont
  {Yoshida}}]{Moller12}%
  \BibitemOpen
  \bibfield  {author} {\bibinfo {author} {\bibfnamefont {P.}~\bibnamefont
  {Moller}}, \bibinfo {author} {\bibfnamefont {W.~D.}\ \bibnamefont {Myers}},
  \bibinfo {author} {\bibfnamefont {H.}~\bibnamefont {Sagawa}}, \ and\ \bibinfo
  {author} {\bibfnamefont {S.}~\bibnamefont {Yoshida}},\ }\href@noop {}
  {\bibfield  {journal} {\bibinfo  {journal} {Phys. Rev. Lett.}\ }\textbf
  {\bibinfo {volume} {108}},\ \bibinfo {pages} {052501} (\bibinfo {year}
  {2012})}\BibitemShut {NoStop}%
\bibitem [{\citenamefont {Zhang}\ and\ \citenamefont {Chen}(2013)}]{Zhang13}%
  \BibitemOpen
  \bibfield  {author} {\bibinfo {author} {\bibfnamefont {Z.}~\bibnamefont
  {Zhang}}\ and\ \bibinfo {author} {\bibfnamefont {L.~W.}\ \bibnamefont
  {Chen}},\ }\href@noop {} {\bibfield  {journal} {\bibinfo  {journal} {Phys.
  Lett. B}\ }\textbf {\bibinfo {volume} {726}},\ \bibinfo {pages} {234}
  (\bibinfo {year} {2013})}\BibitemShut {NoStop}%
\bibitem [{\citenamefont {Hebeler}\ and\ \citenamefont
  {Schwenck}(2014)}]{Heb14}%
  \BibitemOpen
  \bibfield  {author} {\bibinfo {author} {\bibfnamefont {K.}~\bibnamefont
  {Hebeler}}\ and\ \bibinfo {author} {\bibfnamefont {A.}~\bibnamefont
  {Schwenck}},\ }\href@noop {} {\bibfield  {journal} {\bibinfo  {journal} {Eur.
  Phys. J. A}\ }\textbf {\bibinfo {volume} {50}},\ \bibinfo {pages} {11}
  (\bibinfo {year} {2014})}\BibitemShut {NoStop}%
\bibitem [{\citenamefont {Tamii}\ \emph {et~al.}(2014)\citenamefont {Tamii},
  \citenamefont {von Neumann-Cosel},\ and\ \citenamefont
  {Poltoratska}}]{Tam14}%
  \BibitemOpen
  \bibfield  {author} {\bibinfo {author} {\bibfnamefont {A.}~\bibnamefont
  {Tamii}}, \bibinfo {author} {\bibfnamefont {P.}~\bibnamefont {von
  Neumann-Cosel}}, \ and\ \bibinfo {author} {\bibfnamefont {I.}~\bibnamefont
  {Poltoratska}},\ }\href@noop {} {\bibfield  {journal} {\bibinfo  {journal}
  {Eur. Phys. J. A}\ }\textbf {\bibinfo {volume} {50}},\ \bibinfo {pages} {28}
  (\bibinfo {year} {2014})}\BibitemShut {NoStop}%
\bibitem [{\citenamefont {Horowitz}\ and\ \citenamefont
  {Piekarewicz}(2001)}]{Horo01}%
  \BibitemOpen
  \bibfield  {author} {\bibinfo {author} {\bibfnamefont {C.~J.}\ \bibnamefont
  {Horowitz}}\ and\ \bibinfo {author} {\bibfnamefont {J.}~\bibnamefont
  {Piekarewicz}},\ }\href@noop {} {\bibfield  {journal} {\bibinfo  {journal}
  {Phys. Rev. Lett.}\ }\textbf {\bibinfo {volume} {86}},\ \bibinfo {pages}
  {5647} (\bibinfo {year} {2001})}\BibitemShut {NoStop}%
\bibitem [{\citenamefont {Lattimer}\ and\ \citenamefont
  {Prakash}(2001)}]{Lattimer01}%
  \BibitemOpen
  \bibfield  {author} {\bibinfo {author} {\bibfnamefont {J.~M.}\ \bibnamefont
  {Lattimer}}\ and\ \bibinfo {author} {\bibfnamefont {M.}~\bibnamefont
  {Prakash}},\ }\href@noop {} {\bibfield  {journal} {\bibinfo  {journal}
  {Astrophys. J}\ }\textbf {\bibinfo {volume} {550}},\ \bibinfo {pages} {426}
  (\bibinfo {year} {2001})}\BibitemShut {NoStop}%
\bibitem [{\citenamefont {Steiner}\ \emph {et~al.}(2005)\citenamefont
  {Steiner}, \citenamefont {Prakash}, \citenamefont {Lattimer},\ and\
  \citenamefont {Ellis}}]{STEINER05}%
  \BibitemOpen
  \bibfield  {author} {\bibinfo {author} {\bibfnamefont {A.~W.}\ \bibnamefont
  {Steiner}}, \bibinfo {author} {\bibfnamefont {M.}~\bibnamefont {Prakash}},
  \bibinfo {author} {\bibfnamefont {J.~M.}\ \bibnamefont {Lattimer}}, \ and\
  \bibinfo {author} {\bibfnamefont {P.~J.}\ \bibnamefont {Ellis}},\ }\href@noop
  {} {\bibfield  {journal} {\bibinfo  {journal} {Phys. Rep.}\ }\textbf
  {\bibinfo {volume} {411}},\ \bibinfo {pages} {325} (\bibinfo {year}
  {2005})}\BibitemShut {NoStop}%
\bibitem [{\citenamefont {Li}\ and\ \citenamefont {Steiner}(2006)}]{LI06}%
  \BibitemOpen
  \bibfield  {author} {\bibinfo {author} {\bibfnamefont {B.~A.}\ \bibnamefont
  {Li}}\ and\ \bibinfo {author} {\bibfnamefont {A.~W.}\ \bibnamefont
  {Steiner}},\ }\href@noop {} {\bibfield  {journal} {\bibinfo  {journal} {Phys.
  Lett. B}\ }\textbf {\bibinfo {volume} {642}},\ \bibinfo {pages} {436}
  (\bibinfo {year} {2006})}\BibitemShut {NoStop}%
\bibitem [{\citenamefont {Janka}\ \emph {et~al.}(2007)\citenamefont {Janka}
  \emph {et~al.}}]{Jank07}%
  \BibitemOpen
  \bibfield  {author} {\bibinfo {author} {\bibfnamefont {H.-T.}\ \bibnamefont
  {Janka}} \emph {et~al.},\ }\href@noop {} {\bibfield  {journal} {\bibinfo
  {journal} {Phys. Rep.}\ }\textbf {\bibinfo {volume} {442}},\ \bibinfo {pages}
  {38} (\bibinfo {year} {2007})}\BibitemShut {NoStop}%
\bibitem [{\citenamefont {Fischer}\ \emph {et~al.}(2014)\citenamefont {Fischer}
  \emph {et~al.}}]{Fischer14}%
  \BibitemOpen
  \bibfield  {author} {\bibinfo {author} {\bibfnamefont {T.}~\bibnamefont
  {Fischer}} \emph {et~al.},\ }\href@noop {} {\bibfield  {journal} {\bibinfo
  {journal} {Eur. Phys. J. A}\ }\textbf {\bibinfo {volume} {50}},\ \bibinfo
  {pages} {46} (\bibinfo {year} {2014})}\BibitemShut {NoStop}%
\bibitem [{\citenamefont {Lattimer}(2012)}]{Latt12}%
  \BibitemOpen
  \bibfield  {author} {\bibinfo {author} {\bibfnamefont {J.~M.}\ \bibnamefont
  {Lattimer}},\ }\href@noop {} {\bibfield  {journal} {\bibinfo  {journal}
  {Annu. Rev. Nucl. Part. Sci.}\ }\textbf {\bibinfo {volume} {62}},\ \bibinfo
  {pages} {485} (\bibinfo {year} {2012})}\BibitemShut {NoStop}%
\bibitem [{\citenamefont {Steiner}\ and\ \citenamefont
  {Gandolfi}(2012)}]{Stei12}%
  \BibitemOpen
  \bibfield  {author} {\bibinfo {author} {\bibfnamefont {A.~W.}\ \bibnamefont
  {Steiner}}\ and\ \bibinfo {author} {\bibfnamefont {S.}~\bibnamefont
  {Gandolfi}},\ }\href@noop {} {\bibfield  {journal} {\bibinfo  {journal}
  {Phys. Rev. Lett.}\ }\textbf {\bibinfo {volume} {108}},\ \bibinfo {pages}
  {081102} (\bibinfo {year} {2012})}\BibitemShut {NoStop}%
\bibitem [{\citenamefont {Kohley}\ \emph {et~al.}(2013)\citenamefont {Kohley}
  \emph {et~al.}}]{Kohley13}%
  \BibitemOpen
  \bibfield  {author} {\bibinfo {author} {\bibfnamefont {Z.}~\bibnamefont
  {Kohley}} \emph {et~al.},\ }\href@noop {} {\bibfield  {journal} {\bibinfo
  {journal} {Phys. Rev. C}\ }\textbf {\bibinfo {volume} {88}},\ \bibinfo
  {pages} {041601(R)} (\bibinfo {year} {2013})}\BibitemShut {NoStop}%
\bibitem [{\citenamefont {Shetty}\ \emph {et~al.}(2007)\citenamefont {Shetty},
  \citenamefont {Yennello},\ and\ \citenamefont {Souliotis}}]{Shetty07}%
  \BibitemOpen
  \bibfield  {author} {\bibinfo {author} {\bibfnamefont {D.~V.}\ \bibnamefont
  {Shetty}}, \bibinfo {author} {\bibfnamefont {S.~J.}\ \bibnamefont
  {Yennello}}, \ and\ \bibinfo {author} {\bibfnamefont {G.~A.}\ \bibnamefont
  {Souliotis}},\ }\href@noop {} {\bibfield  {journal} {\bibinfo  {journal}
  {Phys. Rev. C}\ }\textbf {\bibinfo {volume} {76}},\ \bibinfo {pages} {024606}
  (\bibinfo {year} {2007})}\BibitemShut {NoStop}%
\bibitem [{\citenamefont {Kohley}\ \emph {et~al.}(2010)\citenamefont {Kohley}
  \emph {et~al.}}]{Kohl10}%
  \BibitemOpen
  \bibfield  {author} {\bibinfo {author} {\bibfnamefont {Z.}~\bibnamefont
  {Kohley}} \emph {et~al.},\ }\href@noop {} {\bibfield  {journal} {\bibinfo
  {journal} {Phys. Rev. C}\ }\textbf {\bibinfo {volume} {82}},\ \bibinfo
  {pages} {064601} (\bibinfo {year} {2010})}\BibitemShut {NoStop}%
\bibitem [{\citenamefont {Lattimer}\ and\ \citenamefont
  {Prakash}(2007)}]{Lat07}%
  \BibitemOpen
  \bibfield  {author} {\bibinfo {author} {\bibfnamefont {J.~M.}\ \bibnamefont
  {Lattimer}}\ and\ \bibinfo {author} {\bibfnamefont {M.}~\bibnamefont
  {Prakash}},\ }\href@noop {} {\bibfield  {journal} {\bibinfo  {journal} {Phys.
  Rep.}\ }\textbf {\bibinfo {volume} {109}},\ \bibinfo {pages} {442} (\bibinfo
  {year} {2007})}\BibitemShut {NoStop}%
\bibitem [{\citenamefont {Sotani}\ \emph {et~al.}(2012)\citenamefont {Sotani},
  \citenamefont {Nakazato}, \citenamefont {Iida},\ and\ \citenamefont
  {Oyamatsu}}]{Sot12}%
  \BibitemOpen
  \bibfield  {author} {\bibinfo {author} {\bibfnamefont {H.}~\bibnamefont
  {Sotani}}, \bibinfo {author} {\bibfnamefont {K.}~\bibnamefont {Nakazato}},
  \bibinfo {author} {\bibfnamefont {K.}~\bibnamefont {Iida}}, \ and\ \bibinfo
  {author} {\bibfnamefont {K.}~\bibnamefont {Oyamatsu}},\ }\href@noop {}
  {\bibfield  {journal} {\bibinfo  {journal} {Phys. Rev. Lett.}\ }\textbf
  {\bibinfo {volume} {108}},\ \bibinfo {pages} {201101} (\bibinfo {year}
  {2012})}\BibitemShut {NoStop}%
\bibitem [{\citenamefont {Steiner}\ \emph {et~al.}(2013)\citenamefont
  {Steiner}, \citenamefont {Lattimer},\ and\ \citenamefont
  {Brown}}]{Steiner_apj13}%
  \BibitemOpen
  \bibfield  {author} {\bibinfo {author} {\bibfnamefont {A.~W.}\ \bibnamefont
  {Steiner}}, \bibinfo {author} {\bibfnamefont {J.~M.}\ \bibnamefont
  {Lattimer}}, \ and\ \bibinfo {author} {\bibfnamefont {E.~F.}\ \bibnamefont
  {Brown}},\ }\href@noop {} {\bibfield  {journal} {\bibinfo  {journal}
  {Astrophys. J. Lett.}\ }\textbf {\bibinfo {volume} {765}},\ \bibinfo {pages}
  {L5} (\bibinfo {year} {2013})}\BibitemShut {NoStop}%
\bibitem [{\citenamefont {Tsang}\ \emph {et~al.}(2009)\citenamefont {Tsang}
  \emph {et~al.}}]{TSANG09}%
  \BibitemOpen
  \bibfield  {author} {\bibinfo {author} {\bibfnamefont {M.~B.}\ \bibnamefont
  {Tsang}} \emph {et~al.},\ }\href@noop {} {\bibfield  {journal} {\bibinfo
  {journal} {Phys. Rev. Lett.}\ }\textbf {\bibinfo {volume} {102}},\ \bibinfo
  {pages} {122701} (\bibinfo {year} {2009})}\BibitemShut {NoStop}%
\bibitem [{\citenamefont {Kohley}\ and\ \citenamefont
  {Yennello}(2014)}]{Kohley14}%
  \BibitemOpen
  \bibfield  {author} {\bibinfo {author} {\bibfnamefont {Z.}~\bibnamefont
  {Kohley}}\ and\ \bibinfo {author} {\bibfnamefont {S.}~\bibnamefont
  {Yennello}},\ }\href@noop {} {\bibfield  {journal} {\bibinfo  {journal} {Eur.
  Phys. J. A}\ }\textbf {\bibinfo {volume} {50}},\ \bibinfo {pages} {31}
  (\bibinfo {year} {2014})}\BibitemShut {NoStop}%
\bibitem [{\citenamefont {Li}\ and\ \citenamefont {Han}(2013)}]{Li13}%
  \BibitemOpen
  \bibfield  {author} {\bibinfo {author} {\bibfnamefont {B.~A.}\ \bibnamefont
  {Li}}\ and\ \bibinfo {author} {\bibfnamefont {X.}~\bibnamefont {Han}},\
  }\href@noop {} {\bibfield  {journal} {\bibinfo  {journal} {Phys. Lett. B}\
  }\textbf {\bibinfo {volume} {727}},\ \bibinfo {pages} {276} (\bibinfo {year}
  {2013})}\BibitemShut {NoStop}%
\bibitem [{\citenamefont {Lattimer}\ and\ \citenamefont
  {Lim}(2013)}]{Lattimer13}%
  \BibitemOpen
  \bibfield  {author} {\bibinfo {author} {\bibfnamefont {J.~M.}\ \bibnamefont
  {Lattimer}}\ and\ \bibinfo {author} {\bibfnamefont {Y.}~\bibnamefont {Lim}},\
  }\href@noop {} {\bibfield  {journal} {\bibinfo  {journal} {Astrophys. J}\
  }\textbf {\bibinfo {volume} {771}},\ \bibinfo {pages} {51} (\bibinfo {year}
  {2013})}\BibitemShut {NoStop}%
\bibitem [{\citenamefont {Vinas}\ \emph {et~al.}(2014)\citenamefont {Vinas},
  \citenamefont {Centelles}, \citenamefont {Roca-Maza},\ and\ \citenamefont
  {Warda}}]{Vin14}%
  \BibitemOpen
  \bibfield  {author} {\bibinfo {author} {\bibfnamefont {X.}~\bibnamefont
  {Vinas}}, \bibinfo {author} {\bibfnamefont {M.}~\bibnamefont {Centelles}},
  \bibinfo {author} {\bibfnamefont {X.}~\bibnamefont {Roca-Maza}}, \ and\
  \bibinfo {author} {\bibfnamefont {M.}~\bibnamefont {Warda}},\ }\href@noop {}
  {\bibfield  {journal} {\bibinfo  {journal} {Eur. Phys. J. A}\ }\textbf
  {\bibinfo {volume} {50}},\ \bibinfo {pages} {27} (\bibinfo {year}
  {2014})}\BibitemShut {NoStop}%
\bibitem [{\citenamefont {McIntosh}\ \emph {et~al.}(2010)\citenamefont
  {McIntosh} \emph {et~al.}}]{MCINTOSH10}%
  \BibitemOpen
  \bibfield  {author} {\bibinfo {author} {\bibfnamefont {A.~B.}\ \bibnamefont
  {McIntosh}} \emph {et~al.},\ }\href@noop {} {\bibfield  {journal} {\bibinfo
  {journal} {Phys. Rev. C}\ }\textbf {\bibinfo {volume} {81}},\ \bibinfo
  {pages} {034603} (\bibinfo {year} {2010})}\BibitemShut {NoStop}%
\bibitem [{\citenamefont {Hudan}\ \emph {et~al.}(2012)\citenamefont {Hudan}
  \emph {et~al.}}]{Hudan12}%
  \BibitemOpen
  \bibfield  {author} {\bibinfo {author} {\bibfnamefont {S.}~\bibnamefont
  {Hudan}} \emph {et~al.},\ }\href@noop {} {\bibfield  {journal} {\bibinfo
  {journal} {Phys. Rev. C}\ }\textbf {\bibinfo {volume} {86}},\ \bibinfo
  {pages} {021603(R)} (\bibinfo {year} {2012})}\BibitemShut {NoStop}%
\bibitem [{\citenamefont {Brown}\ \emph {et~al.}(2013)\citenamefont {Brown}
  \emph {et~al.}}]{Brown13}%
  \BibitemOpen
  \bibfield  {author} {\bibinfo {author} {\bibfnamefont {K.}~\bibnamefont
  {Brown}} \emph {et~al.},\ }\href@noop {} {\bibfield  {journal} {\bibinfo
  {journal} {Phys. Rev. C}\ }\textbf {\bibinfo {volume} {87}},\ \bibinfo
  {pages} {061601} (\bibinfo {year} {2013})}\BibitemShut {NoStop}%
\bibitem [{\citenamefont {Hudan}\ and\ \citenamefont
  {deSouza}(2014)}]{Hudan14}%
  \BibitemOpen
  \bibfield  {author} {\bibinfo {author} {\bibfnamefont {S.}~\bibnamefont
  {Hudan}}\ and\ \bibinfo {author} {\bibfnamefont {R.~T.}\ \bibnamefont
  {deSouza}},\ }\href@noop {} {\bibfield  {journal} {\bibinfo  {journal} {Eur.
  Phys. J. A}\ }\textbf {\bibinfo {volume} {50}},\ \bibinfo {pages} {36}
  (\bibinfo {year} {2014})}\BibitemShut {NoStop}%
\bibitem [{\citenamefont {Bocage}\ \emph {et~al.}(2000)\citenamefont {Bocage}
  \emph {et~al.}}]{Boc00}%
  \BibitemOpen
  \bibfield  {author} {\bibinfo {author} {\bibfnamefont {F.}~\bibnamefont
  {Bocage}} \emph {et~al.},\ }\href@noop {} {\bibfield  {journal} {\bibinfo
  {journal} {Nucl. Phys. A}\ }\textbf {\bibinfo {volume} {676}},\ \bibinfo
  {pages} {391} (\bibinfo {year} {2000})}\BibitemShut {NoStop}%
\bibitem [{\citenamefont {Davin}\ \emph {et~al.}(2002)\citenamefont {Davin}
  \emph {et~al.}}]{Davin02}%
  \BibitemOpen
  \bibfield  {author} {\bibinfo {author} {\bibfnamefont {B.}~\bibnamefont
  {Davin}} \emph {et~al.},\ }\href@noop {} {\bibfield  {journal} {\bibinfo
  {journal} {Phys. Rev. C}\ }\textbf {\bibinfo {volume} {65}},\ \bibinfo
  {pages} {064614} (\bibinfo {year} {2002})}\BibitemShut {NoStop}%
\bibitem [{\citenamefont {Colin}\ \emph {et~al.}(2003)\citenamefont {Colin}
  \emph {et~al.}}]{Colin03}%
  \BibitemOpen
  \bibfield  {author} {\bibinfo {author} {\bibfnamefont {J.}~\bibnamefont
  {Colin}} \emph {et~al.},\ }\href@noop {} {\bibfield  {journal} {\bibinfo
  {journal} {Phys. Rev. C.}\ }\textbf {\bibinfo {volume} {67}},\ \bibinfo
  {pages} {064603} (\bibinfo {year} {2003})}\BibitemShut {NoStop}%
\bibitem [{\citenamefont {Piantelli}\ \emph {et~al.}(2002)\citenamefont
  {Piantelli} \emph {et~al.}}]{Pia02}%
  \BibitemOpen
  \bibfield  {author} {\bibinfo {author} {\bibfnamefont {S.}~\bibnamefont
  {Piantelli}} \emph {et~al.},\ }\href@noop {} {\bibfield  {journal} {\bibinfo
  {journal} {Phys. Rev. Lett.}\ }\textbf {\bibinfo {volume} {88}},\ \bibinfo
  {pages} {052701} (\bibinfo {year} {2002})}\BibitemShut {NoStop}%
\bibitem [{\citenamefont {Filippo}\ \emph
  {et~al.}(2005{\natexlab{a}})\citenamefont {Filippo} \emph {et~al.}}]{Fil05}%
  \BibitemOpen
  \bibfield  {author} {\bibinfo {author} {\bibfnamefont {E.~D.}\ \bibnamefont
  {Filippo}} \emph {et~al.},\ }\href@noop {} {\bibfield  {journal} {\bibinfo
  {journal} {Phys. Rev. C}\ }\textbf {\bibinfo {volume} {71}},\ \bibinfo
  {pages} {044602} (\bibinfo {year} {2005}{\natexlab{a}})}\BibitemShut
  {NoStop}%
\bibitem [{\citenamefont {Filippo}\ and\ \citenamefont {Pagano}(2014)}]{Fil14}%
  \BibitemOpen
  \bibfield  {author} {\bibinfo {author} {\bibfnamefont {E.~D.}\ \bibnamefont
  {Filippo}}\ and\ \bibinfo {author} {\bibfnamefont {A.}~\bibnamefont
  {Pagano}},\ }\href@noop {} {\bibfield  {journal} {\bibinfo  {journal} {Eur.
  Phys. J. A}\ }\textbf {\bibinfo {volume} {50}},\ \bibinfo {pages} {32}
  (\bibinfo {year} {2014})}\BibitemShut {NoStop}%
\bibitem [{\citenamefont {Papa}\ \emph {et~al.}(2005)\citenamefont {Papa},
  \citenamefont {Giuliani},\ and\ \citenamefont {Bonasera}}]{Papa05}%
  \BibitemOpen
  \bibfield  {author} {\bibinfo {author} {\bibfnamefont {M.}~\bibnamefont
  {Papa}}, \bibinfo {author} {\bibfnamefont {G.}~\bibnamefont {Giuliani}}, \
  and\ \bibinfo {author} {\bibfnamefont {A.}~\bibnamefont {Bonasera}},\
  }\href@noop {} {\bibfield  {journal} {\bibinfo  {journal} {J. Comput. Phys.}\
  }\textbf {\bibinfo {volume} {208}},\ \bibinfo {pages} {403} (\bibinfo {year}
  {2005})}\BibitemShut {NoStop}%
\bibitem [{\citenamefont {Filippo}\ \emph
  {et~al.}(2005{\natexlab{b}})\citenamefont {Filippo} \emph {et~al.}}]{Fil05b}%
  \BibitemOpen
  \bibfield  {author} {\bibinfo {author} {\bibfnamefont {E.~D.}\ \bibnamefont
  {Filippo}} \emph {et~al.},\ }\href@noop {} {\bibfield  {journal} {\bibinfo
  {journal} {Phys. Rev. C}\ }\textbf {\bibinfo {volume} {71}},\ \bibinfo
  {pages} {064604} (\bibinfo {year} {2005}{\natexlab{b}})}\BibitemShut
  {NoStop}%
\bibitem [{\citenamefont {Russotto}\ \emph {et~al.}(2010)\citenamefont
  {Russotto} \emph {et~al.}}]{Rus10}%
  \BibitemOpen
  \bibfield  {author} {\bibinfo {author} {\bibfnamefont {P.}~\bibnamefont
  {Russotto}} \emph {et~al.},\ }\href@noop {} {\bibfield  {journal} {\bibinfo
  {journal} {Phys. Rev. C}\ }\textbf {\bibinfo {volume} {81}},\ \bibinfo
  {pages} {064605} (\bibinfo {year} {2010})}\BibitemShut {NoStop}%
\bibitem [{\citenamefont {Steiner}\ and\ \citenamefont
  {Li}(2005)}]{Steiner_05}%
  \BibitemOpen
  \bibfield  {author} {\bibinfo {author} {\bibfnamefont {A.~W.}\ \bibnamefont
  {Steiner}}\ and\ \bibinfo {author} {\bibfnamefont {B.~A.}\ \bibnamefont
  {Li}},\ }\href@noop {} {\bibfield  {journal} {\bibinfo  {journal} {Phys. Rev.
  C}\ }\textbf {\bibinfo {volume} {72}},\ \bibinfo {pages} {041601(R)}
  (\bibinfo {year} {2005})}\BibitemShut {NoStop}%
\bibitem [{\citenamefont {Tsang}\ \emph {et~al.}(2004)\citenamefont {Tsang}
  \emph {et~al.}}]{TSANG04}%
  \BibitemOpen
  \bibfield  {author} {\bibinfo {author} {\bibfnamefont {M.~B.}\ \bibnamefont
  {Tsang}} \emph {et~al.},\ }\href@noop {} {\bibfield  {journal} {\bibinfo
  {journal} {Phys. Rev. Lett.}\ }\textbf {\bibinfo {volume} {92}},\ \bibinfo
  {pages} {062701} (\bibinfo {year} {2004})}\BibitemShut {NoStop}%
\bibitem [{\citenamefont {Papa}\ \emph {et~al.}(2001)\citenamefont {Papa},
  \citenamefont {Maruyama},\ and\ \citenamefont {Bonasera}}]{PAPA01}%
  \BibitemOpen
  \bibfield  {author} {\bibinfo {author} {\bibfnamefont {M.}~\bibnamefont
  {Papa}}, \bibinfo {author} {\bibfnamefont {T.}~\bibnamefont {Maruyama}}, \
  and\ \bibinfo {author} {\bibfnamefont {A.}~\bibnamefont {Bonasera}},\
  }\href@noop {} {\bibfield  {journal} {\bibinfo  {journal} {Phys. Rev. C}\
  }\textbf {\bibinfo {volume} {64}},\ \bibinfo {pages} {024612} (\bibinfo
  {year} {2001})}\BibitemShut {NoStop}%
\bibitem [{\citenamefont {Papa}\ \emph {et~al.}(2007)\citenamefont {Papa} \emph
  {et~al.}}]{Papa07}%
  \BibitemOpen
  \bibfield  {author} {\bibinfo {author} {\bibfnamefont {M.}~\bibnamefont
  {Papa}} \emph {et~al.},\ }\href@noop {} {\bibfield  {journal} {\bibinfo
  {journal} {Phys. Rev. C}\ }\textbf {\bibinfo {volume} {75}},\ \bibinfo
  {pages} {054616} (\bibinfo {year} {2007})}\BibitemShut {NoStop}%
\bibitem [{\citenamefont {Amorini}\ \emph {et~al.}(2009)\citenamefont {Amorini}
  \emph {et~al.}}]{Amo09}%
  \BibitemOpen
  \bibfield  {author} {\bibinfo {author} {\bibfnamefont {F.}~\bibnamefont
  {Amorini}} \emph {et~al.},\ }\href@noop {} {\bibfield  {journal} {\bibinfo
  {journal} {Phys. Rev. Lett.}\ }\textbf {\bibinfo {volume} {102}},\ \bibinfo
  {pages} {112701} (\bibinfo {year} {2009})}\BibitemShut {NoStop}%
\bibitem [{\citenamefont {Cardella}\ \emph {et~al.}(2012)\citenamefont
  {Cardella} \emph {et~al.}}]{Card12}%
  \BibitemOpen
  \bibfield  {author} {\bibinfo {author} {\bibfnamefont {G.}~\bibnamefont
  {Cardella}} \emph {et~al.},\ }\href@noop {} {\bibfield  {journal} {\bibinfo
  {journal} {Phys. Rev. C}\ }\textbf {\bibinfo {volume} {85}},\ \bibinfo
  {pages} {064609} (\bibinfo {year} {2012})}\BibitemShut {NoStop}%
\end{thebibliography}

%

\end{document}